\documentclass[aps,10pt,twocolumn,superscriptaddress]{revtex4-2}
\usepackage[colorlinks=true,allcolors=blue]{hyperref}
\usepackage{amssymb}
\usepackage{amsmath}
\usepackage{graphicx}
\usepackage[caption=false]{subfig}
\usepackage{amsfonts}
\usepackage{xcolor}
\usepackage{epsfig}
\usepackage{color}
\usepackage{bm}
\usepackage{tabularx}
\usepackage{multirow}
\usepackage{mathtools}
\usepackage{xfrac}
\usepackage{blkarray}
\usepackage{bbold}
\usepackage{float}
\usepackage[mathscr]{euscript}
\usepackage{autobreak}
\usepackage{comment}
\usepackage[utf8]{inputenc}
\usepackage{newunicodechar} % Define custom Unicode characters
\newunicodechar{̈}{\\"} % Map U+0308 to umlaut command
\newcommand{\vb}{V$_{\rm B}$}
\newcommand{\cn}{C$_{\rm N}$}
\newcommand{\cb}{C$_{\rm B}$}
\newcommand{\on}{O$_{\rm N}$}

\usepackage{soul}
\graphicspath{{./Figures/}}

\begin{document}
\title{Interacting donor-acceptor pairs as the origin of coupled spin-optical signals in hexagonal boron nitride}

\author{Guanjian Hu}%
\affiliation{Beijing Computational Science Research Center, Beijing 100193, China
}%

\author{Jijun Huang}%
\affiliation{Beijing Computational Science Research Center, Beijing 100193, China
}%

\author{Bing Huang}%
\email{bing.huang@csrc.ac.cn}
\affiliation{Beijing Computational Science Research Center, Beijing 100193, China
}%
\affiliation{School of Physics and Astronomy, Beijing Normal University, Beijing 100875, China}

\author{Song Li}%
\email{li.song@csrc.ac.cn}
\affiliation{Beijing Computational Science Research Center, Beijing 100193, China
}%

\date{\today}

\begin{abstract}
Optically addressable spin defects in hexagonal boron nitride hold promise for room-temperature quantum technologies, but their microscopic identities remain largely unknown. Using first-principles calculations, we show that coupled spin-optical signals arise from interacting donor–acceptor pairs, not the commonly believed isolated defects. Intra- and inter-pair separations control charge transfer, electronic structure, and spin coupling, thereby greatly modulating zero-phonon lines, phonon sidebands, lifetimes, and the sign of optically detected magnetic resonance contrast. Importantly, we identify two distinct charge-state-dependent coupling regimes and extend this picture to correlated defect ensembles, explaining the wide diversity of experimental observations. Our results establish a microscopic framework for coupled defect behavior and provide design principles for spin-active quantum emitters in wide-bandgap semiconductors.
\end{abstract}

\maketitle

%
%
%----------INTRODUCTION----------%

$\textit{Introduction}$---Color centers are point defects in crystals whose spin degrees of freedom could be coupled to optical transitions. Color centers in wide bandgap semiconductors are promising building blocks for room-temperature quantum information processing~\cite{wolfowicz2021quantum,de2021materials}. Among them, the nitrogen-vacancy (NV) center in diamond is a prototypical system that has been extensively studied and employed in quantum sensing and quantum networks~\cite{dolde2011electric,maze2008nanoscale,balasubramanian2008nanoscale,lesik2019magnetic,maurer2012room,fuchs2011quantum}. Despite its success, the near-surface NV center suffers from surface noise, which degrades the charge stability and shortens the coherence time~\cite{rosskopf2014investigation,sangtawesin2019origins}. 
In the past years, hexagonal boron nitride (hBN), a layered material, has emerged as a promising alternative host platform~\cite{sajid2020single,gilardoni2026optically,tran2016quantum,gottscholl2020initialization,chejanovsky2021single,mendelson2021identifying}. Compared to bulk materials, hBN offers improved compatibility and integrability with other quantum photonic architectures, such as waveguides and optical cavities~\cite{caldwell2019photonics,montblanch2023layered}. Moreover, color centers in hBN can be deterministically fabricated in proximity to the hBN surface, enabling high scalability~\cite{xu2021creating}.

The negatively charged boron vacancy (\vb)~\cite{ivady2020ab} has a triplet ground state and has been coherently controlled as a quantum sensor~\cite{gottscholl2020initialization,Haykal2022,mu2022excited,vaidya2023quantum,liu2022coherent}. However, its weak brightness and broad phonon sideband (PSB) limit optical performance. In contrast, single photon emitters (SPEs) in hBN often display sharp zero phonon lines (ZPLs) and high brightness~\cite{tran2016robust,bourrellier2016bright,guo2023coherent,stern2024quantum}. The microscopic origins of these emissions remain under active investigation due to the difficulty of direct identification, motivating the use of theoretical calculation and simulation for defect assignment. In addition to native defects~\cite{weston2018native,li2025native,huang2012defect}, defect structures with carbon~\cite{mackoit2019carbon,li2022carbon,jara2021first,li2022bistable,ortigoza2022thermodynamic,babar2025carbon,benedek2023symmetric,li2022ultraviolet}, oxygen~\cite{li2023prolonged,li2022identification,cholsuk2025raman}, and hydrogen~\cite{turiansky2019dangling} impurities involved have been widely proposed as candidates for SPEs. Carbon are commonly present in hBN synthesized by various methods~\cite{mendelson2021identifying,chejanovsky2021single,pelini2019shallow,plo2025isotope,iwanski2024revealing,gale2022site,bourrellier2016bright,tang2025structured} and carbon-related defects have been linked to optical-active centers, while clusters such as dimer~\cite{mackoit2019carbon}, trimer~\cite{jara2021first,li2022carbon}, tetramers~\cite{benedek2023symmetric} and chain complexes~\cite{maciaszek2024blue} reproduce key optical signatures--such as ZPL, PSB,  fluorescence lifetime--observed in experiments. 

Although substantial evidence strongly indicates that carbon-related defects play a key role in SPEs spanning from the visible to ultraviolet range, the similarity of their optical signatures makes it difficult to provide a one-to-one correspondence between specific defect structure and individual emission lines. More importantly, most existing interpretations rely on isolated-defect models, whereas real materials typically host multiple defects nearby. Coupled defects, which are common in doped hBN samples~\cite{tan2022donor,mejia2024general}, can have obvious interactions that modify charge states, induce charge transfer, and introduce additional optical and spin transition pathways. To address this problem, in this work, we go beyond the single-defect picture and investigate the coupled defects in hBN using first-principles calculations. We show that inter-defect separation influences charge transfer, electronic structure, and spin interactions. We further examine how such pairwise interactions can be extended to more complex multi-defect environments, providing a basis for understanding experimentally observed coupled spin systems.

$\textit{First-principles calculations}$---We performe density functional theory (DFT) calculations using the Vienna ab initio simulation package (VASP)~\cite{kresse1996efficiency, kresse1996efficient}. A plane-wave cutoff energy of 400 eV is selected. The projector augmented wave (PAW) potentials are adopted to describe the properties of valence electrons and core regions~\cite{blochl1994projector, kresse1999ultrasoft}. The geometry optimization is done with the generalized gradient approximation of Perdew, Burke, and Ernzerhof (PBE). A correction for the
van der Waals interactions is included within the Grimme-D3 scheme~\cite{grimme2010consistent}. The electronic structures of defects are calculated with the Heyd–Scuseria–Ernzerhof (HSE) hybrid density functional~\cite{heyd2003hybrid}. A three-dimensional $6\times6\times2$ supercell model is employed to explore the possible separations between defect pairs. The $\Gamma$-point sampling scheme was used for all calculations, and the convergence threshold for atomic forces is set to $10^{-2}$ eV/\AA. The $\Delta$SCF method is adopted to evaluate the electronic excited states~\cite{gali2008ab}. In addition, energy correction is added to the calculated ZPL~\cite{mackoit2019carbon}. Defect formation energy is calculated with charge correction~\cite{freysoldt2009fully}. More calculation details about the radiative lifetime, zero field splitting (ZFS) are in Supplementary Materials (SM) Sec. I. QuTiP~\cite{johansson2012qutip} is used to simulate the spin and photodynamics, as described in SM Sec. II.

\begin{figure}\label{figure1}
    \centering
    \includegraphics[width=1\linewidth]{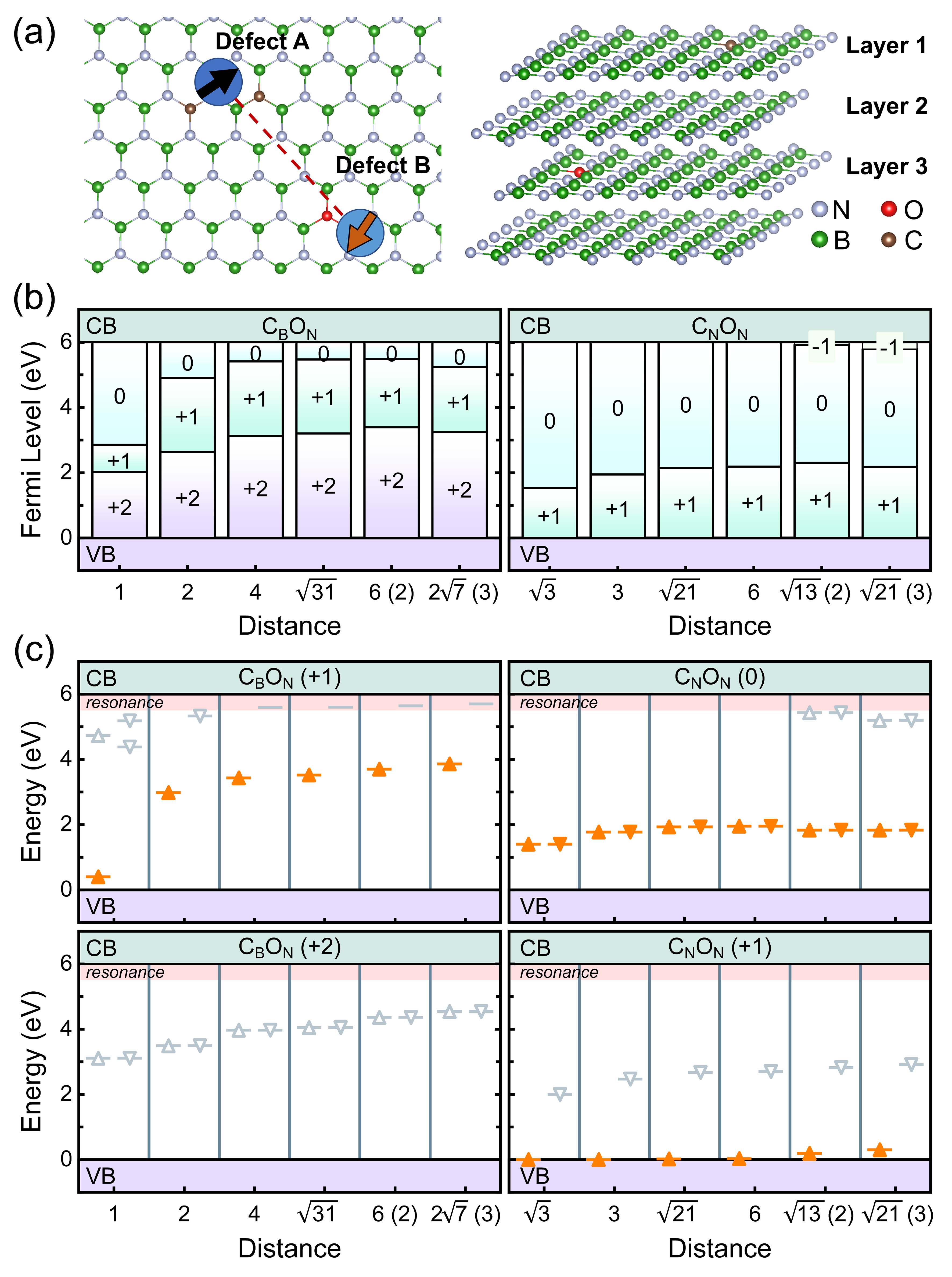}
    \caption{(a) Structure of coupled spin defects A and B in hBN. A 4-layer model is used in this paper and the defects can reside on either the same or distinct layer of hBN. (b) The CTLs of \cb-\on\ and \cn-\on\ pairs. The distances are defined according to their separation in the ideal hBN lattice, with “1” being the B-N bond. (c) The defect energy level diagrams of \cb-\on\ and \cn-\on\ pairs at different charge states. The orange and gray colors indicate the occupied and empty states. The shaded area is the resonant state due to the wavefunction hybridization of \on\ and CBM. Numbers in parentheses in the $x$-axis denote the layer number of the second defect.}
    \label{figure1}
\end{figure}

$\textit{Carbon-oxygen defect pairs models}$---We first consider the carbon-oxygen defect pairs as minimal models to elucidate how defect separation influences the optical and spin properties of coupled defects. These models serve as the most straightforward starting point for establishing the key physics underlying the more general defect-pair picture discussed below. Both \cb, \cn, and \on\ can be readily incorporated under typical growth conditions~\cite{maciaszek2022thermodynamics}. We focus on \cb-\on\ and \cn-\on, while the \cn-\cb\ has been discussed previously~\cite{auburger2021towards}. A four-layer supercell model is used to construct both in-plane and interlayer defect configurations [Fig.~\ref{figure1}(a)]. 
%The calculated formation energies and charge transition levels (CTLs) are shown in Fig. S1 and Fig.~\ref{figure1}(b). For \cb-\on, the stable charge states in gap are $q$ = +2, +1, 0. Since both \cb\ and \on\ are donor-like defects, the formation energy of \cb-\on\ is relatively high. 

\begin{figure}[H]
	\centering
	\includegraphics[width=1\linewidth]{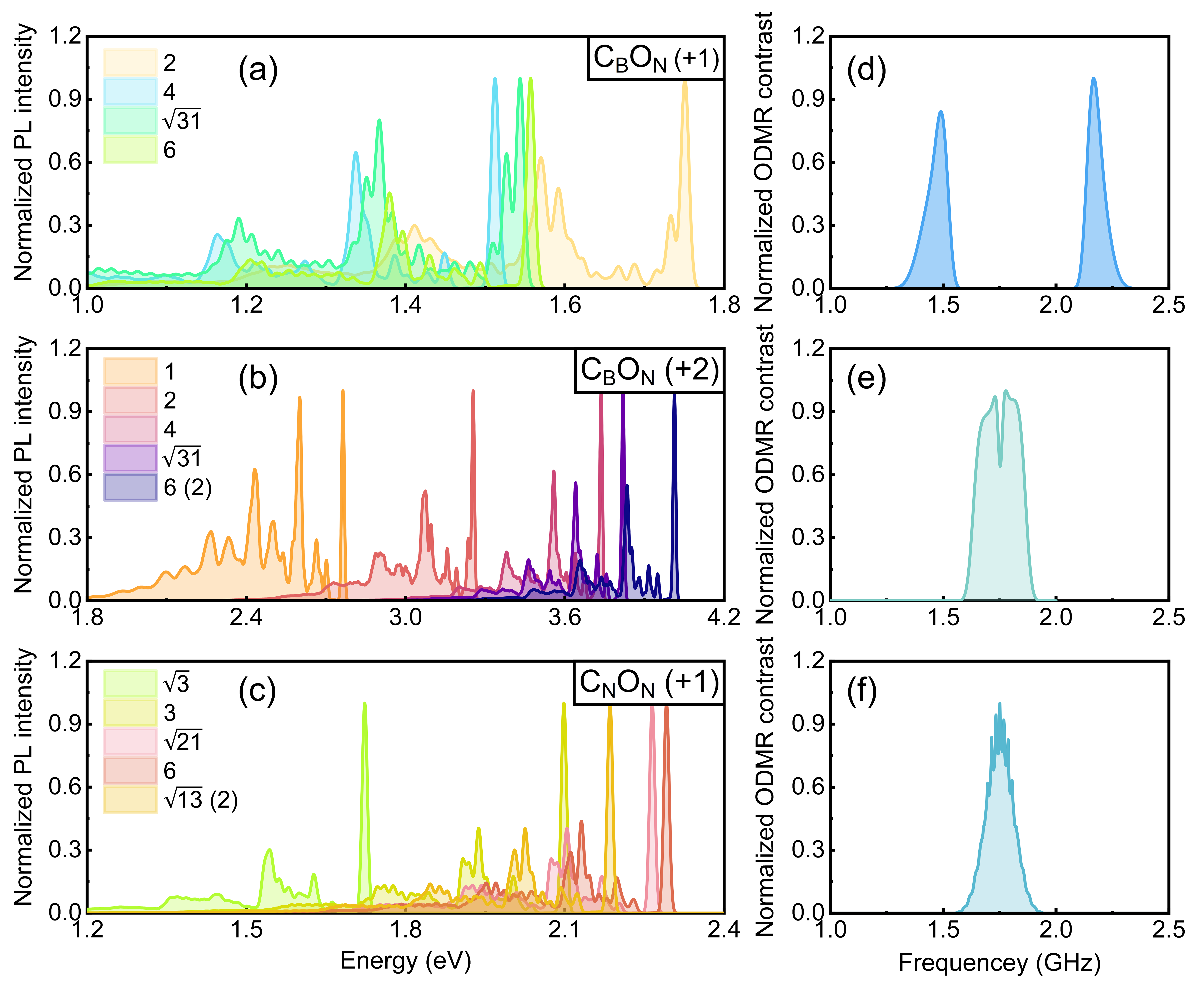}
	\caption{(a-c) The simulated PSB of \cb-\on\ and \cn-\on\ pairs in different charge states. The energy is aligned to ZPLs. Numbers in parentheses denote the layer number of the second defect. The simulated ODMR spectra of (d) \cb\on(+), (e) \cb-\on-$\sqrt{31}$(+),  and (f) \cn-\on(+) pairs with $^{13}\text{C}$. The magnetic field is set to 62.5 mT.}
	\label{figure2}
\end{figure}
 
For the nearest-neighbor \cb\on\ configuration, strong Coulomb repulsion drives an out-of-plane distortion, indicating the absence of a stable chemical bond [SM Fig. S2]. This instability is confirmed by phonon calculation, which shows one imaginary phonon mode with an energy of 70.4 meV in $C_{2v}$ configuration. 
%With optimized out-of-plane distortion, the C-O distance is 2.21 \AA\ at neutral charge state, remarkably longer than the typical C-O bond. The corresponding wavefunction exhibits antibonding character, indicating the absence of a true chemical bond. 
The stable charge states in gap are $q$ = +2 and 0 with the singlet ground state, while the $q$ = +1 state is only stable within a narrow Fermi energy range (2.09 to 2.83 eV). The hyperfine constant $A_{zz}$ of $^{13}\text{C}$ is 842 MHz [SM Tab. S1], which is significantly larger than the hyperfine splitting in group-III defects observed~\cite{gao2025single}, suggesting it is unlikely to account for the coherently controlled nuclear spin. 
%At the neutral charge state, the occupied state originates from the dangling bond of $p_x$ orbital, while the unoccupied state comes from the out-of-plane $p_z$  orbital of the C atom. 
In the $q = +2$ charge state, the defect retrieves the planar $C_{2v}$ symmetry, and there is no occupied defect levels in gap. 
%The defect levels in the gap are from the out-of-plane orbitals of the C atom and neighboring B atoms, while the $p_x$ dangling bond becomes saturated. 
The optical transition occurs between the valence band edge (VBM) and the $p_z$ orbital of the carbon atom. The ZPL energy is 2.76 eV with Huang-Rhys (HR) factor $S_{\text{HR}}$ = 2.57, making it a candidate for nonmagnetic blue emitters.

As the separation distance between \cb\ and \on\ increases to \cb-\on-2, the Coulomb interaction is reduced, leading to significant changes in charge transition levels (CTLs) and optical properties. We note strong hybridization between the defect level of \on\ and the conduction band edge, leading to resonant and quasilocalized states near the conduction band minimum (CBM). At $q = +1$ charge state, the reduced energy difference between defect levels enables charge-transfer-like optical transitions, resulting in ZPL energies in the visible range (1.75--1.45 eV) and relatively small HR factor (1.39--2.11 eV) [SM Tab. S3], in excellent agreement with experimental data~\cite{tran2016robust}.
%positions shift and the stable charge states alter to $q$ = +1 and +2. Since \on\ is a a stronger donor, the CTL $\epsilon$(+1/0) of \cb-\on\ converges to that of \on\ and the CTL $\epsilon$(+2/+1) of \cb-\on\ converges to the CTL $\epsilon$(+1/0) of \cb\ with increasing separation. The calculated ZPL energies and HR factors are summarized in Tab. S3. We also find the defect level of \on\ strongly hybridizes with the conduction band edge, leading to resonant and quasilocalized states near the conduction band minimum (CBM), as shown in Fig.~\ref{figure1}(c). At +1 charge state, the energy difference between defect levels of \cb\ and \on\ decreases due to reduced Coulomb repulsion, resulting in lower ZPL energies ranging from 1.75 to 1.45 eV in the visible region. The corresponding HR factors of these structures range from 2.11 to 1.39, in excellent agreement with experimental data~\cite{tran2016robust}. 
To gain further insight into the vibrational properties, we calculate the phonon spectral function. The corresponding PSBs are dominated by high-energy localized modes around the defect ($\sim$ 180 meV), with additional contributions arising from interlayer vibrations when the defects reside in different layers [SM Fig. S3]. This phenomenon highlights that the single-layer hBN model is insufficient to fully capture vibrational effects from interlayer coupling, especially when the wavefunctions of defects have extended out-of-plane distributions. In contrast, at $q = +2$ charge state, the optical transition is from VBM to the defect level of \cb, and the ZPL increases with separation, eventually converging to the ultraviolet region at 4.1 eV.

We next turn to \cn-\on, which forms a typical donor-acceptor pair (DAP). At neutral charge state, charge transfer from \on\ to \cn\ stabilizes a nonmagnetic ground state. Compared to \cb-\on, the donor level of \on\ is shifted deeper into the gap due to Coulomb attraction. As the separation increases, the ZPL of neutral \cn-\on\ decreases from the ultraviolet to the blue region, while it increases from 1.72 eV to 2.19 eV at $q = +1$ state [SM Tab. S3]. The result indicates that defect-defect coupling could significantly modify the optical properties of individual emitters, providing an alternative explanation for the broad distribution of emission lines in experiments besides the local strain effects~\cite{grosso2017tunable}.

%with a stable nonmagnetic state at $q$ = 0 where an electron on \on\ transferred to \cn. The CTL $\epsilon$(+1/0) is at 1.53 eV. As obtained earlier, the concentration of \cn-\on\ shall be high due to the low formation energy when O and C are present under N-poor conditions~\cite{maciaszek2022thermodynamics}. Compared to \cb-\on, the defect level of \on\ in \cn-\on\ DAPs lies deeper in gap due to the attraction between the donor and acceptor. The calculated ZPL is 4.01 eV, together with HR factor of 1.89. As the separation increases, the ZPL decreases to 2.94 eV, as summarized in Tab. S4. In contrast, at +1 charge state, the ZPL increases from 1.72 eV to 2.19 eV as the separation increases. In all DAPs configurations, the optical excitation is from carbon substitution, and the PSB features remain similar.

Based on the hyperfine matrix, we simulate the cw-ODMR spectra and directly compare them with experiments. Inclusion of $^{13}$C hyperfine coupling leads to pronounced peak splitting and asymmetric ODMR lineshapes due to anisotropic interactions [Fig.~\ref{figure2}(d-e)]. The distorted \cb\on\ configurations exhibit linewidth broadening around 69 MHz, much wider than the 16 MHz in the experiment~\cite{gao2025single}. The unequal intensities of the ODMR peaks originate from the anisotropic hyperfine interaction, which induces mixing of the spin states. As a result, the transition probabilities are no longer identical, leading to an asymmetric spectral profile. In contrast, \cn-\on\ DAPs have smaller A$_{zz}$ of $^{13}$C on \cn, which leads to merged ODMR peaks.

%The hyperfine parameters are evaluated for the most abundant nuclear spin-active isotopes ($^{11}$B, $^{14}$N, $^{13}$C). Based on previous studies, it suffices to only include the atoms up to second neighbors, and the nuclear Zeeman and quadrupole terms can be safely neglected~\cite{auburger2021towards}. We first use the ODMR spectrum of \cb\ as a benchmark, which has a full width at half maximum (FWHM) of 43 MHz. Our simulated hyperfine matrix of \cb\ is slightly underestimated due to the reduced supercell size, resulting in ODMR line broadening around 38 MHz. Inclusion of $^{13}$C hyperfine coupling leads to pronounced peak splitting and asymmetric ODMR lineshapes due to anisotropic interactions, as shown in Fig.~\ref{figure2}(d-e). For distorted \cb\on, the linewidth broadening is around 69 MHz, much wider than the 16 MHz in the experiment~\cite{gao2025single}. The unequal intensities of the ODMR peaks originate from the anisotropic hyperfine interaction, which induces mixing of the spin states. As a result, the transition probabilities are no longer identical, leading to an asymmetric spectral profile. Even at large separations, coupling between defects remains observable. \cn-\on\ DAPs exhibit broader FWHM linewidths due to contributions from neighboring boron atoms, while the smaller A$_{zz}$ of $^{13}$C on \cn\ leads to merged ODMR peaks.

\begin{figure}
	\centering
	\includegraphics[width=1\linewidth]{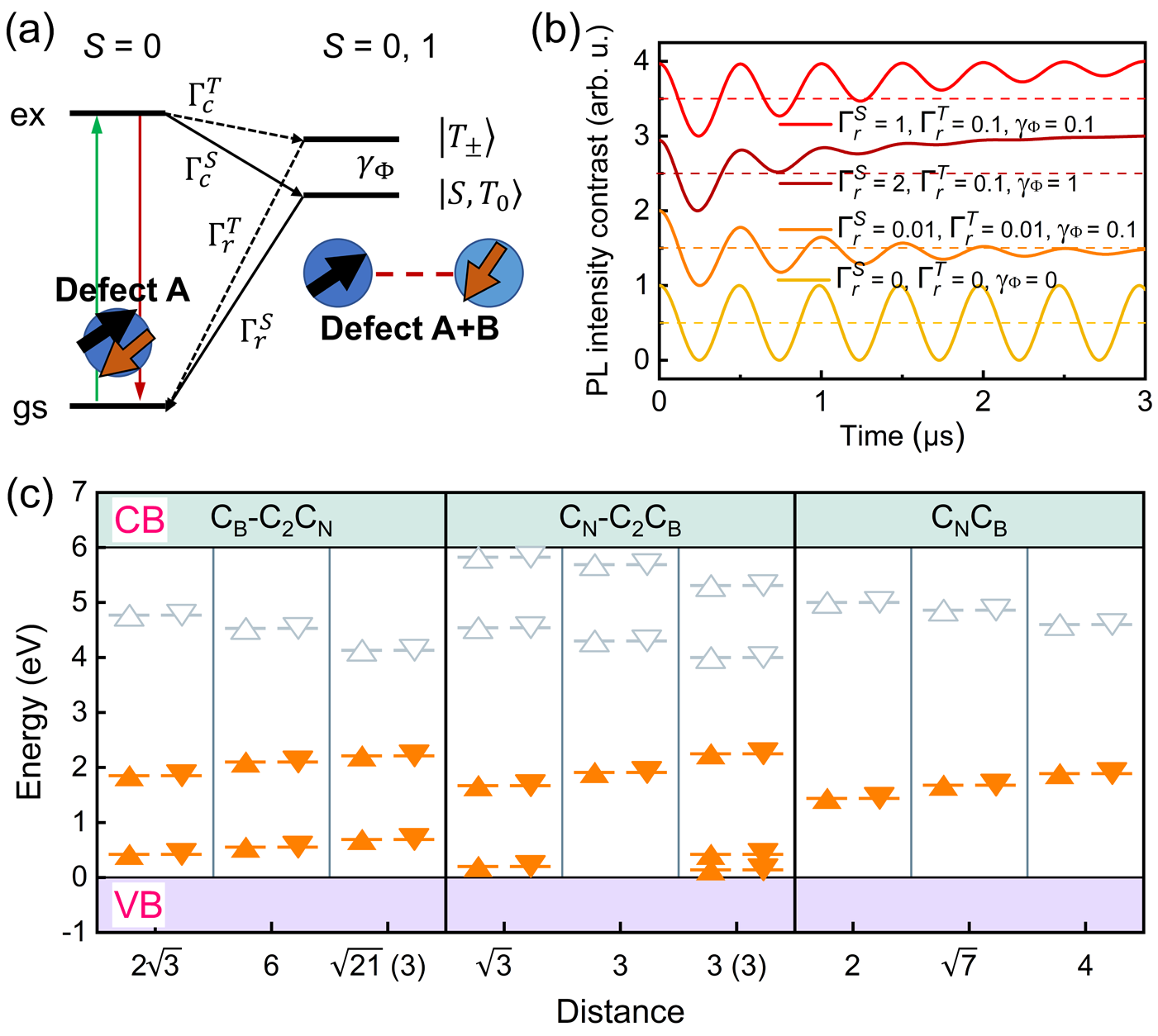}
	\caption{(a) The coupled spin model of defect A and B. The bright emission is from the ground state to the local excited state when two electrons are located on defect A in a singlet state. On the right, the electrons are on different defects due to charge transfer and have $S$ = 0, 1 states. $\Omega$ is the Rabi driving frequency and $\gamma_{\Phi}$ is the transition rate between $\left|S,T_{0}\right\rangle$ and $\left|T_{\pm}\right\rangle$. Black solid and dashed lines indicate the allowed and forbidden transitions. (b) The simulated ODMR contrast from Rabi oscillation based on the various parameters from the model. (c) Possible candidates for defect A and the corresponding defect energy level diagrams in the neutral charge state. Numbers in parentheses denote the layer number of the second defect. }
    \label{figure3}
\end{figure}

$\textit{Spin and photodynamics}$---We next propose that neutral \cn-\on\ DAPs could serve as a model for the recently observed optical-spin defect pairs (OSDP)~\cite{singh2025violet,robertson2025charge,whitefield2026narrowband} [Fig.~\ref{figure3}(a)]. In this model, two electrons can either localize on a single defect (acceptor-like \cn), forming a singlet ground state, or occupy separate defects, giving rise to both singlet and triplet configurations. These many-body states can be divided into two groups: one is the $m_s$ = 0 singlet-triplet mixed configurations $\left|\downarrow\uparrow\right\rangle$ $\pm$ $\left|\uparrow\downarrow\right\rangle$ (denoted as $\left|S,T_{0}\right\rangle$) and the other one is the $m_s$ = $\pm$1 pure triplet states ($\left|T_{\pm}\right\rangle$). The two groups are split by ZFS and external magnetic field $g\mu_{B}B_0$. The relative ordering and splitting of these states depend sensitively on the spatial separation between defects, reflecting the spatial anisotropy of the spin dipole-dipole interaction when the wavefunction distribution of the two electrons extends in hBN plane. With current configurations, the ZFS is at hundreds of MHz [SM Tab. S5]. At large distances (several nm), the ZFS is expected to decrease to negligible, and the external magnetic field dominates the energy splitting. The behavior may explain the singular resonance peak on ODMR in the previous experiment. 
%The calculated ZFS parameters are listed in Tab. S5. \cn-\on\/-XX has ZFS $\textbf{D}$ = 258 MHz with $\textbf{E}$ = -47.5 MHz so the $\left|T_{\pm}\right\rangle$ lies higher than $\left|S,T_{0}\right\rangle$. The ZFS alters its sign as the separation increases due to the spatial anisotropy of the spin dipole-dipole interaction when the wavefunction distribution of the two electrons extends in hBN plane. At large distances (several nm), the ZFS is expected to decrease to negligible, and the external magnetic field dominates the energy splitting. The behavior may explain the singular resonance peak on ODMR in the previous experiment. These results demonstrate that even the simplest coupled defect pairs already exhibit pronounced separation-dependent behavior in their CTLs, optical excitation, and spin properties. 

Fig.~\ref{figure3}(a) presents the schematic energy level diagram of the DAPs. The above-mentioned many-body states lie close in energy to the spin-conserving excited state. The optical transition from the singlet ground state to $\left|S,T_{0}\right\rangle$ is allowed, while that to $\left|T_{\pm}\right\rangle$ is forbidden without spin-orbit coupling. Since the $\Gamma^{S}_c$ and $\Gamma^{S}_r$ are generally faster than $\Gamma^{T}_c$ and $\Gamma^{T}_r$, respectively, the population dynamics depend on the relative rates. If $\Gamma^{S}_c$/$\Gamma^{S}_r$ $>$ $\Gamma^{T}_c$/$\Gamma^{T}_r$ ($\Gamma^{S}_c$/$\Gamma^{S}_r$ $<$ $\Gamma^{T}_c$/$\Gamma^{T}_r$), then the $\left|S,T_{0}\right\rangle$ ($\left|T_{\pm}\right\rangle$) state will have higher population under optical pumping and alter the sign of ODMR contrast. Fig.~\ref{figure3}(b) shows the simulated ODMR contrast based on the coupled spin pair model with different sets of $\Gamma^{S}_r$, $\Gamma^{T}_r$, and $\gamma_{\Phi}$, indicating various kinds of decay curves in the Rabi oscillations. These results demonstrate that even the simplest defect pairs exhibit strong separation-dependent behavior in their CTLs, ZPLs, and spin dynamics. This picture connects the optical signatures of DAPs with their spin-dependent response, establishing coupled defect pairs as the fundamental units of OSDPs.
%We could roughly estimate the rate based on the radiative lifetime. $\Gamma^{S}_c$ is 10$^{4}$ faster than $\Gamma^{S}_r$ due to large wavefunction overlap and optical transition dipole moment, and decreases as separation increases. Note that the nonradiative process and local metastable triplet could provide additional pathways to influence the dynamics~\cite{whitefield2026narrowband}. 

\begin{figure}
	\centering
	\includegraphics[width=1\linewidth]{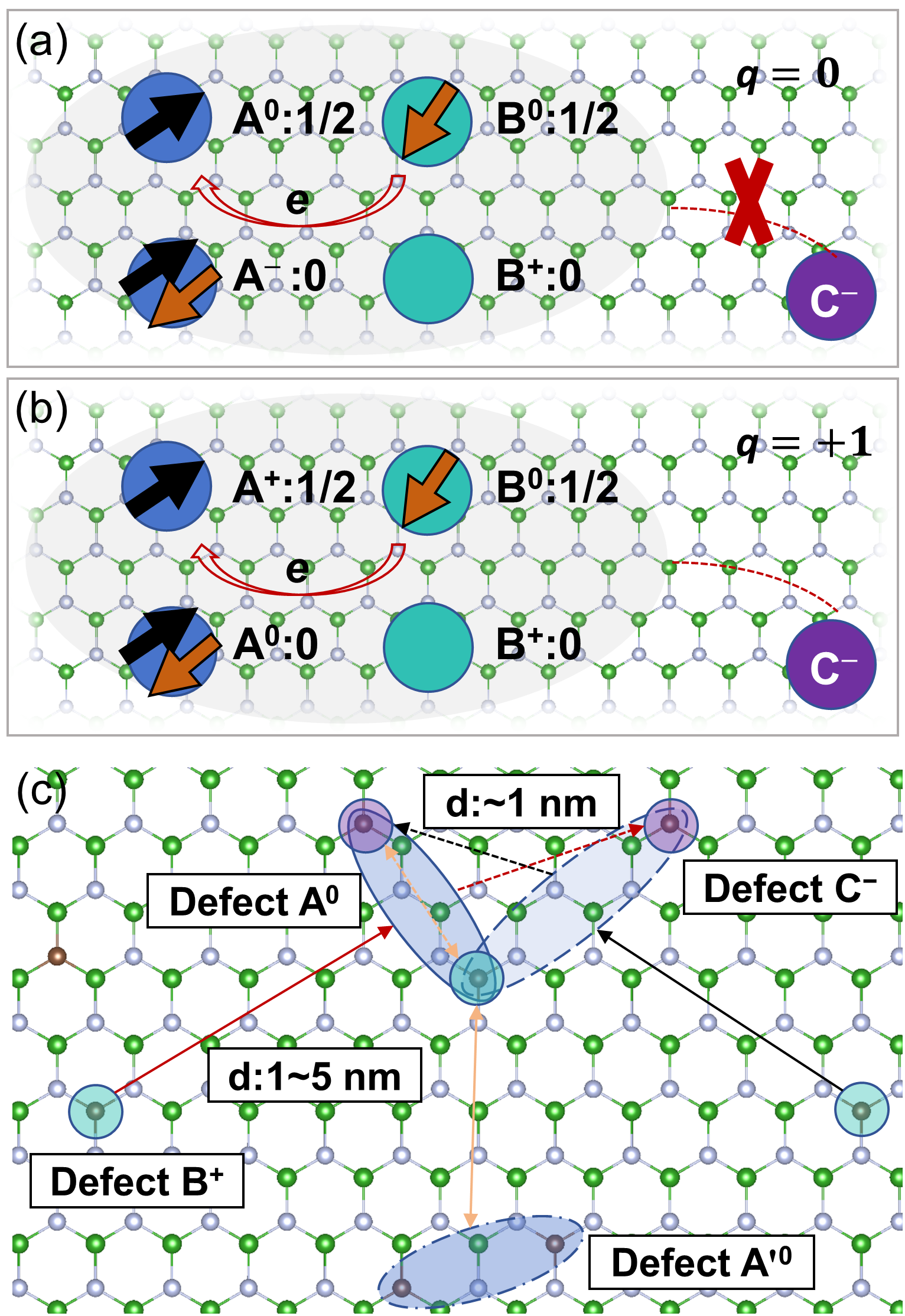}
	\caption{(a-b) Two representative charge states of weakly coupled spin system. In $q$ = 0 condition, the total charge state of defect A and B is neutralized, typically occurs in DAPs, and charge transfer occurs only within the system; the $q$ = +1 configuration typically corresponds to defects without clear donor or acceptor character, and requires defect C as compensation. (c) Correlated-defects ensemble picture. Two scenarios exist with simple \cb\ and \cn\ defects as examples. In the first one, defect A acts as a \cn-\cb\ DAP and exchanges charge with a nearby compensating defect C$^-$ (\cn) while also interacting with donor-like defect B$^+$ (\cb), so that the DAP is effectively neutralized as A$^0$. In the second one, defect A interacts with another DAP, A$^{\prime}$, such that charge transfer occurs between the constituent donor and acceptor defects of the two pairs, leading to two charge-neutral coupled units, A$^0$ and A$^{\prime 0}$.}
    \label{figure4}
\end{figure}

$\textit{Coupled spin defects}$---Experimentally, the wide variation in creation and recombination rates in the ensemble suggests the presence of multiple types of coupled spin systems~\cite{robertson2025charge}. While isolated DAPs already capture key features of charge transfer and spin interactions, they are insufficient to explain the full range of observed behaviors.
%while neutral \cb-\on\ could also form such systems, the energy of the local excited state is also close to the energy of charge transfer states due to the strong hybridization between \on\ and CBM. Better localization of defect levels in gap could improve the brightness and contrast, for example in \vbon-\cb\ from our previous study~\cite{li2025quantum} and \cb-\cb\ ~\cite{du2025carbon}. 
Although isolated DAPs of simple substitutional defects are charge-neutralized [Fig.~\ref{figure4}(a)], a general limitation is that they typically provide only a single defect level [see Fig.~\ref{figure1}(c)], so optical excitation necessarily involves the band edges, resulting in relatively long excited-state lifetimes in microseconds~\cite{vaidya2025coherent,liu2025experimental}. Naturally, we can consider a minimal coupled system consisting of a DAP interacting with an additional nearby defect~\cite{robertson2025charge}. In this picture, the DAP serves as the primary optical center (namely defect A), providing localized occupied and unoccupied states that enable bright local emission, while the additional defect (namely defect B) introduces spin degrees of freedom through weak coupling. Both the optical activity from the DAP and spin response from inter-defect coupling offer a simple mechanism to reconcile bright emission with spin-dependent signals.

%Such coupled configurations also relax the limitations of simple substitutional defects, which typically host only a single defect level and therefore rely on band-edge transitions. In contrast, defect complexes can support multiple localized states within the gap, leading to shorter radiative lifetimes and enhanced optical contrast. The resulting electronic structure is highly sensitive to charge transfer and defect separation, providing a natural explanation for the variability of ZPLs and ODMR responses observed experimentally.

However, most DAPs in hBN demonstrate a singlet ground state at neutral charge state over a large range of the band gap. Therefore, the S = 1/2 spin state requires the DAPs to be positively charged. The stabilization of these coupled spin systems generally requires charge compensation, implying the presence of additional nearby defects (defect C in Fig.~\ref{figure4}(b)). This leads to a more realistic picture in which DAPs are embedded in a correlated defect environment [Fig.~\ref{figure4}(c)]. As illustrated by the red arrows, defect A can be charged by a nearby compensating defect C$^{-}$ and subsequently neutralized through charge transfer from defect B. Similar DAP may also form in blue dashed circle and interact with another set of defects B and C, as indicated by the black arrows.

Beyond such isolated DAP-defect interactions, we further propose that DAP-DAP coupling can occur, giving rise to a network of charge transfer and spin interaction pathways. In this scenario, defect A remains neutral while coupling to defect B, which is stabilized in a positive charge state and forms an effective coupled center A$^{\prime}$ with a nearby compensating defect C, as illustrated by the orange arrows. Within such an ensemble, both intra-pair separation and inter-pair coupling govern the optical and spin properties. As a result, the observed ODMR signals arise not from isolated defects but from interacting DAPs forming a network of coupled spin centers. This framework accounts for the broad distribution of ZPLs, PSBs, and ODMR contrasts, and can be extended to defect systems beyond carbon in hBN as well as to other defects ensembles in wide-bandgap semiconductors.

$\textit{Experimental signatures and validation}$---To connect our model with experiments, we focus on defects consistent with the observed narrow ODMR linewidths, which point to carbon-related centers. Other acceptors, e. g. \vb-$n$H~\cite{weston2018native,maciaszek2026cbvb}, may also contribute, but the ODMR linewidth would be large and their roles require further experimental verification. In the proposed framework, Defect B should be donor-like, with simple candidates including \cb\ and C$_2$\cb\ if defect complexes are excluded. If defect A acts as an acceptor, possible carbon substitutional defects include C$_2$\cn, C$_4$\cn, and more generally the odd-number carbon clusters with surplus \cn. However, at negative charge state, these acceptors usually exhibit ZPL in the blue to violet region ($>$ 3 eV). We speculate that these pure carbon acceptors are unlikely to explain the visible emission lines. To solve this, we propose defect A itself is a DAP, such as \cn-\cb, C$_2$\cn-\cb, or C$_2$\cb-\cn, as energy level diagrams shown in Fig.~\ref{figure3}(c). In these complexes, the two defects are separated by just several lattice constants (i. e. $<$ 1 nm), which is energetically favorable due to Coulomb attraction. At such short separations, defect A exhibits no obvious donor or acceptor character, so it couples only weakly with donor defect B. The ZPLs range from 1.74 to 2.73 eV with HR factor around 2$-$3 [SM Tab. S6], in good agreement with experimental observations. Under N-poor conditions, the positive charge state of defect A, as well as the associated charge transfer process, can be stabilized, where the Fermi level is pinned near VBM + 1.8 eV~\cite{maciaszek2022thermodynamics}. Simultaneously, the concentration of acceptors would be higher, further facilitating the charging of the coupled spin system. Such a carbon-defects ensemble provides a comprehensive explanation for the visible-to-ultraviolet bright emission with photodynamics due to coupled spins.

$\textit{Conclusions}$---In summary, we demonstrate that inter-defect coupling play a central role in determining the optical and spin properties of defects in hBN. Using carbon–oxygen pairs as minimal models, we identify DAPs as the key units of coupled spin systems and reveal two distinct coupling configurations governed by charge transfer. We extend the coupled-spin model from the binary defects to a general multi-defect framework, and demonstrate that coupled spin behavior must be understood based on interacting DAPs within a correlated defects ensemble. The intra-pair and inter-pair interactions collectively govern the charge transfer, spin dynamics, and optical responses. This framework resolves the origin of ambiguities in the microscopic interpretation of coupled spin defects in hBN, and can be extended to other wide bandgap semiconductors. These insights offer practical guidance for the rational design of quantum defects and qubits toward scalable quantum information processing.

$\textit{Competing interests}$---The authors declare that there are no competing interests.

%
%
%----------ACKNOWLEDGEMENT----------%
$\textit{Acknowledgements}$---This work is supported by Science Challenge Project (Grant No. TZ2025013). B.H. acknowledges the NSFC (Grant No.\ W2511008, 12088101), the National Key Research and Development of China (Grant No.\ 2022YFA1402400), and NSAF (Grant No.\ U2230402). 

$\textit{Data Availability}$---The data that support the findings of this study are available from the corresponding author upon reasonable request.

\bibliography{name.bib}
%\begin{thebibliography}{99}  

\end{document}